\documentclass[12pt,preprint]{aastex}  
\usepackage{mathptmx}
\usepackage{courier}
\usepackage{graphicx}
\usepackage{colordvi}
\usepackage{natbib}
\usepackage{color}

\newcommand{\CaII}{\ion{Ca}{2}}

\newcommand{\kms}{km$\,$s$^{-1}$}
\newcommand{\arcpix}{$^{\prime\prime}$~pixel$^{-1}$}
\newcommand{\ha}{H$\alpha$}

\shorttitle{\ha\ Imaging Spectroscopy of a Circular Flare's Remote Ribbon}
\shortauthors{Deng et al.}

\begin{document}

\title{High-Cadence and High-Resolution \ha\ Imaging Spectroscopy of a Circular Flare's Remote Ribbon with IBIS}

\author{NA DENG\altaffilmark{1}, ALEXANDRA TRITSCHLER\altaffilmark{2}, JU JING\altaffilmark{1}, XIN CHEN\altaffilmark{1}, CHANG LIU\altaffilmark{1}, KEVIN REARDON\altaffilmark{2,3,4}, CARSTEN DENKER\altaffilmark{5}, YAN XU\altaffilmark{1}, AND HAIMIN WANG\altaffilmark{1}}
\affil{1. Space Weather Research Laboratory, New Jersey Institute of Technology, University Heights, Newark, NJ 07102-1982, USA; na.deng@njit.edu}
\affil{2. National Solar Observatory, Sacramento Peak, Sunspot, NM 88349-0062, USA;}
\affil{3. INAF - Osservatorio Astrofisico di Arcetri, Largo E. Fermi 5, 50125 Florence, Italy;}
\affil{4. Astrophysics Research Centre, Queen's University, Belfast, BT7 1NN, Northern Ireland, UK;}
\affil{5. Leibniz-Institut f\"ur Astrophysik Potsdam, An der Sternwarte 16, 14482 Potsdam, Germany}

\begin{abstract}
We present an unprecedented high-resolution \ha\ imaging spectroscopic observation of a C4.1 flare taken with the Interferometric Bidimensional Spectrometer (IBIS) in conjunction with the adaptive optics system at the 76~cm Dunn Solar Telescope on 2011 October 22 in active region NOAA 11324. Such a two-dimensional spectroscopic observation covering the entire evolution of a flare ribbon with high spatial (0.1\arcpix\ image scale), cadence (4.8~s) and spectral (0.1~\AA\ stepsize) resolution is rarely reported. The flare consists of a main circular ribbon that occurred in a parasitic magnetic configuration and a remote ribbon that was observed by the IBIS. Such a circular-ribbon flare with a remote brightening is predicted in 3D fan-spine reconnection but so far has been rarely observed.

During the flare impulsive phase, we define ``core'' and ``halo'' structures in the observed ribbon based on IBIS narrowband images in the \ha\ line wing and line center. Examining the \ha\ emission spectra averaged in the flare core and halo areas, we find that only those from the flare cores show typical nonthermal electron beam heating characteristics that have been revealed by previous theoretical simulations and observations of flaring \ha\ line profiles. These characteristics include: broad and centrally reversed emission spectra, excess emission in the red wing with regard to the blue wing (i.e., red asymmetry), and redshifted bisectors of the emission spectra. We also observe rather quick timescales for the heating ($\sim$30~s) and cooling ($\sim$14--33~s) in the flare core locations. Therefore, we suggest that the flare cores revealed by IBIS track the sites of electron beam precipitation with exceptional spatial and temporal resolution. The flare cores show two-stage motion (a parallel motion along the ribbon followed by an expansion motion perpendicular to the ribbon) during the two impulsive phases of the flare. Some cores jump quickly (30 \kms) between discrete magnetic elements implying reconnection involving different flux tubes. We observe a very high temporal correlation ($\gtrsim0.9$) between the integrated \ha\ and HXR emission during the flare impulsive phase. A short time delay (4.6~s) is also found in the \ha\ emission spikes relative to HXR bursts. The ionization timescale of the cool chromosphere and the extra time taken for the electrons to travel to the remote ribbon site may contribute to this delay.

\end{abstract}

\keywords{
    Sun: flares ---
    Sun: atmospheric motions ---
    Sun: chromosphere ---
    Sun: magnetic fields ---
    line: profiles}

\section{INTRODUCTION}\label{sec:introduction}

Flares are three-dimensional (3D) phenomena involving a variety of physical processes such as magnetic reconnection, energy release, particle acceleration, plasma heating and mass motion. These processes give rise to an enhanced emission at various wavelengths throughout the solar spectrum covering different atmospheric layers, which give us opportunities to scrutinize the flare phenomena from different perspectives \citep[see reviews by, e.g.,][]{Benz2008LRSP....5....1B, Fletcher+etal2011SSRv..159...19F}. For the longest time (since the 1930s), the strong \ha\ line has been an important diagnostic tool to examine the signature of flares in the chromosphere \citep[e.g.,][and references therein]{Hudson2007ASPC..368..365H}. A flare often shows bright expanding ribbons or patches in \ha\ images \citep[e.g.,][]{Zirin1988assu.book.....Z}. The \ha\ emission in those bright ribbons could be due to various chromospheric heating mechanisms during flares. The impulsive heating of \ha\ flare kernels is predominantly due to precipitation of energetic (nonthermal) electron beams \citep[e.g.,][]{Canfield+etal1990ApJ...363..318C, RubioCosta2011PhDT........11R} that are accelerated in the coronal reconnection region and responsible for the hard X-rays (HXR) emission at the flare loop footpoints via collisional thick-target bremsstrahlung \citep{Brown1971SoPh...18..489B, LinR+Hudson1976SoPh...50..153L, Emslie1978ApJ...224..241E}. The chromospheric brightening can also be heated by thermal conduction from the overlying flared hot coronal loops \citep{Zarro+Lemen1988ApJ...329..456Z, Czaykowska+A+D2001ApJ...552..849C, Battaglia+F+B2009A&A...498..891B}, or indirectly by soft
X-rays (SXR) and EUV irradiation emitted from heated coronal and transition region plasmas \citep{Somov1975SoPh...42..235S, Henoux+Nakagawa1977A&A....57..105H, Hawley+Fisher1992ApJS...78..565H, Allred+etal2005ApJ...630..573A}.

Superior to \ha\ filtergram observations that only reveal dynamic morphology and intensity, the complete \ha\ spectra contain much more diagnostic capabilities. The observed \ha\ spectra of flares usually show a variety of shapes and spectral signatures that can be compared with theoretical calculations of \ha\ line profiles based on various physical conditions of the chromosphere during flares. For example, \citet{Canfield+G+R1984ApJ...282..296C}
compute the \ha\ profiles for modeled flare chromospheres in response to three specific excitation processes: nonthermal electron beam precipitation, heat conduction, and high coronal pressure. They find that \ha\ spectra respond sensitively to these processes. In particular, a central reversal with broad enhanced wings is a remarkable signature of \ha\ line profile when the chromosphere is bombarded by an intense flux of nonthermal electrons. The broadened and centrally reversed \ha\ emission profile is also linked to nonthermal effects by precipitating energetic electrons in other numerical simulations of flare chromospheres \citep{Canfield+Gayley1987ApJ...322..999C, Fang+etal1993A&A...274..917F}.

The \ha\ emission profiles have been observed to show such theoretically predicted enhanced broad wing and central reversal signatures in flare kernels that are spatially and temporally well correlated with HXR emissions during the impulsive phase of flares \citep{Canfield+Gunkler1985ApJ...288..353C, LeeS+etal1996ApJ...470L..65L}, which justified the nonthermal electron beam heating mechanism in those areas and the characteristic \ha\ line profile corresponding to such heating mechanism. Based on this remarkable spectral characteristics, the observed impulsive-phase \ha\ emission profiles have been used as a diagnostic tool to identify the sites of nonthermal electron beam precipitation into the chromosphere and to compare them with HXR emissions that originate from the same effect \citep{Canfield+etal1990ApJ...363..318C, Canfield+etal1993ApJ...411..362C}. 

It is noted, however, other effects and heating mechanisms can also produce centrally reversed \ha\ line profiles \citep[e.g.,][]{Abbett+Hawley1999ApJ...521..906A, Kasparova+etal2009A&A...499..923K, RubioCosta2011PhDT........11R}. Therefore, this characteristic profile shape itself can not solely determine the heating mechanism. Nevertheless, the close temporal correlation between variations of HXR flux and \ha\ emission (with short time lag in \ha\ emission) and fast heating and cooling time for the chromosphere can add strong evidence to the nonthermal electron beam heating mechanism \citep{Heinzel1991SoPh..135...65H, Kasparova+etal2009A&A...499..923K}. The existence of nonthermal heating processes does not exclude other heating mechanisms. In many cases, different heating mechanisms work together during flares \citep{Cheng+D+L2006ApJ...653..733C}.

Besides the characteristic shapes, the shift and asymmetry of the line profile can also reveal dynamic phenomena of flare-related mass motions. During the impulsive phase of flares, redshifted \ha\ emission spectra along with conspicuous red asymmetry (i.e., excess red wing emission and the line wing bisectors shift more toward red) are most commonly observed in small bright flare kernels \citep{TangF1983SoPh...83...15T, Ichimoto+Kurokawa1984SoPh...93..105I} although blueshifts and blue asymmetry are occasionally found in the early impulsive phase \citep{Canfield+etal1990ApJ...363..318C}. The redshifts and red asymmetry of \ha\ emission spectra are found closely associated with intense HXR or microwave emissions in space and time \citep{Wuelser+Marti1989ApJ...341.1088W, Canfield+etal1990ApJ...363..318C}. Those redshifted and red asymmetric \ha\ emissions are attributed to downward moving chromospheric condensation \citep{Canfield+etal1990ApJ...348..333C, Gan+R+F1993ApJ...416..886G, Ding+Fang1996SoPh..166..437D}, which is driven by explosive chromospheric evaporation caused by impulsive heating of the upper chromosphere due to bombardment of energetic electrons \citep{Fisher+C+M1985ApJ...289..434F, Fisher+C+M1985ApJ...289..425F, Fisher+C+M1985ApJ...289..414F}. Therefore, the redshifts and red asymmetry could be additional characteristics for \ha\ emission spectra during nonthermal electron beam heating.

Compared to extensive \ha\ imaging observations, two-dimensional (2D) \ha\ spectral observations that spatially and temporally cover a flare are relatively scarce. The aforementioned observations of 2D \ha\ spectra during flares were mainly acquired by traditional one-dimensional (1D) single-slit scanning spectrographs that usually have low spatial resolution ($> 2^{\prime\prime}$) and low cadence (e.g., $>$ 10~s for a small region)  due to the time consumed to scan a 2D region with the 1D slit \citep[e.g.,][]{Canfield+etal1990ApJ...363..318C, Wuelser+etal1994ApJ...424..459W, Ding+F+H1995SoPh..158...81D}. Recently, modern 2D imaging spectroscopic instruments based on tunable narrowband filters (e.g., Fabry-P\'erot etalons) are routinely available and capable of high-spatial (sub arcsecond) and high-temporal (a few seconds) resolution spectroscopy or spectropolarimetry, for example, the Interferometric Bidimensional Spectrometer \citep[IBIS;][]{Cavallini2006SoPh..236..415C}. They are especially useful for observing small-scale or highly dynamic phenomena, such as the chromosphere, jets, and flares \citep[e.g.,][]{Cauzzi+etal2008A&A...480..515C, Cauzzi+etal2009A&A...503..577C}. However, flare spectra taken by these kinds of instruments are rarely published so far. \citet{Kleint2012ApJ...748..138K} reported a spectropolarimetric observation of a C-class flare taken with IBIS during only the decaying phase of the flare. Moving or suddenly appeared brightenings inside the flare ribbons are observed to show strong emission in the chromospheric \CaII\ 8542~\AA\ line profiles. \citet{SanchezAndrade+etal2008A&A...486..577S} exhibited a few \ha\ spectra of two sympathetic mini-flares observed by the G\"ottingen Fabry-P\'erot spectrometer. In the present paper, we will present an unprecedented 2D \ha\ spectroscopic observation carried out by IBIS covering the morphology and entire evolution of a flare ribbon with high spatial (0.1\arcpix\ image scale), temporal (4.8~s), and spectral (0.1~\AA\ sampling stepsize) resolution, which provides detailed morphological, dynamic, and spectral diagnostics of the flare.

It is noteworthy that 3D spectroscopy (i.e., the $x$, $y$, $\lambda$ data cube is simultaneously captured in one exposure although with limited field-of-view and spectral sampling) provided by Multichannel Subtractive Double Pass (MSDP) imaging spectrograph \citep{MeinP1991A&A...248..669M, MeinP2002A&A...381..271M} has played an important role in very high temporal resolution (sub second) studies of \ha\ spectra of flare kernels as well as the spatial and temporal correlation between \ha\ and HXR emissions \citep{Radziszewski+etal2006AdSpR..37.1317R, Radziszewski+R+P2007A&A...461..303R, Radziszewski+R+P2011A&A...535A.123R}.

While the classical 2D flare model \citep{Carmichael1964NASSP..50..451C, Sturrock1966Natur.211..695S, Hirayama1974SoPh...34..323H, Kopp+Pneuman1976SoPh...50...85K} can explain most of the well defined two-ribbon flares, some observed flares  differ from the typical two-ribbon ones, such as \textsf{J}-shaped ribbon flares, multi-ribbon flares, etc. Recent 3D numerical simulations that involve 3D coronal null point reconnection do expect flares having a closed circular ribbon associated with a jet or a remote brightening occurring in a fan-spine magnetic topology \citep{Masson+etal2009ApJ...700..559M, Pariat+etal2010ApJ...714.1762P}. However, observations of such circular-ribbon flares with a jet or a remote bright ribbon are very rare and have only been reported in a few papers \citep{Ugarte-Urra+W+W2007ApJ...662.1293U, Masson+etal2009ApJ...700..559M, Reid+etal2012A&A, Wang+Liu2012ApJ...760..101W, LiuC+etal2013ApJ}. The present paper adds another example of a confined circular-ribbon flare with a concurrent remote bright ribbon (SOL2011-10-22T15:20(C4.1)).

\section{OBSERVATIONS}\label{sec:observation}

We carried out an observing campaign to study chromospheric jets from 2011 October 17 to October 23 in National Solar Observatory/Sacramento Peak using the IBIS instrument at the 76~cm Dunn Solar Telescope that is equipped with high-order adaptive optics (AO) system \citep{Rimmele+Marino2011LRSP....8....2R}. This ground-based observation was coordinated with space-based observation by the Hinode satellite \citep{Kosugi+etal2007SoPh..243....3K}. On October 22, we targeted active region NOAA 11324 (N11$^\circ$ E18$^\circ$) and captured the remote ribbon of a C4.1 flare SOL2011-10-22T15:20 from onset to decay during a good seeing period.

IBIS consists of two synchronized channels: a narrowband (0.022 \AA\ FWHM around 6563~\AA) channel with two tunable Fabry-P\'erot interferometers (FPI) to take 2D spectral data and a broadband ($\sim100$~\AA) channel to take white-light reference images (at 6600~\AA\ in our case) for calibration and post-facto image processing purpose \citep{Cavallini2006SoPh..236..415C, Reardon+Cavallini2008A&A...481..897R}. An \ha\ prefilter of 2.6 \AA\ FWHM centered around 6562.8 \AA\ was used in the narrowband channel to isolate one of the periodic transmission peaks of the FPIs. The IBIS with a round field-of-view (FOV) of 90$^{\prime\prime}$ diameter and 0.1\arcpix\ detector image scale repetitively scanned the H$\alpha$ line from 6561.1 to 6563.8~\AA~(i.e., from $-$1.7 to $+$1.0 \AA\ about the \ha\ line center 6562.8~\AA) using 28 equidistant steps of 0.1~\AA\ stepsize. The observing campaign was aimed at studying high-speed upflowing jets, so we sampled more wavelength points toward the blue wing. Each spectral scan took about 4.8~s with a frame rate of $\sim$6 frames per second and an exposure time of 30~ms for each frame. The time lapse between two consecutive wavelength points is about 0.17~s. The observed ribbon was close to the AO lock point (a small pore) in the FOV of IBIS, giving excellent image quality and stability. The flare was not that strong (C4.1), so the detector of the IBIS narrowband channel for \ha\ spectra acquisition was never saturated throughout the observation.

The Solar Optical Telescope \citep[SOT;][]{Tsuneta+etal2008SoPh..249..167T} of Hinode targeted the same region during the same time period. High resolution (0.16\arcpix, 64~s cadence) Na\,\textsc{i}\,D$_1$ 5896 \AA\ line-of-sight (LOS) magnetograms were recorded by the Narrowband Filter Imager (NFI) to provide detailed information of the magnetic structure.

The temporal and spatial relationship between the \ha\
and HXR emission during the flares impulsive phase can provide important clues on energy transport mechanisms. The evolution of the flare HXR emission was entirely registered by the Reuven Ramaty High Energy Solar Spectroscopic Imager \citep[RHESSI;][]{LinR+etalRHESSI2002SoPh..210....3L}. CLEAN images \citep{HurfordG+etal2002SoPh..210...61H} in the nonthermal energy range (12--25 keV) were reconstructed using the front segments of
detectors 2--8 with 48 s integration time throughout the event. The cadence of RHESSI X-ray light curves is 4~s, which is similar to the cadence (4.8~s) of our \ha\ line scans.

To provide the observational context for the entire flare region, we also use LOS magnetograms taken by the Helioseismic and Magnetic Imager \citep[HMI;][]{Schou+etal2012SoPh..275..229S} and images taken by the Atmospheric Imaging Assembly \citep[AIA;][]{Lemen+etal2012SoPh..275...17L} aboard the Solar Dynamics Observatory (SDO).

\section{DATA REDUCTION}\label{sec:reduction}

We first performed the standard calibration procedures for IBIS observations, which include dark and flat-field correction, alignment and destretch of images in each spectral scan using broadband white-light images as a reference, and correction of blue shifts across the FOV because of the collimated mount of IBIS FPIs. We then aligned a 45-minute time sequence consisting of 550 spectral scans that cover the entire flare duration, again using broadband white-light images as a reference.

\begin{figure}[t]
  \epsscale{0.5}
  \plotone{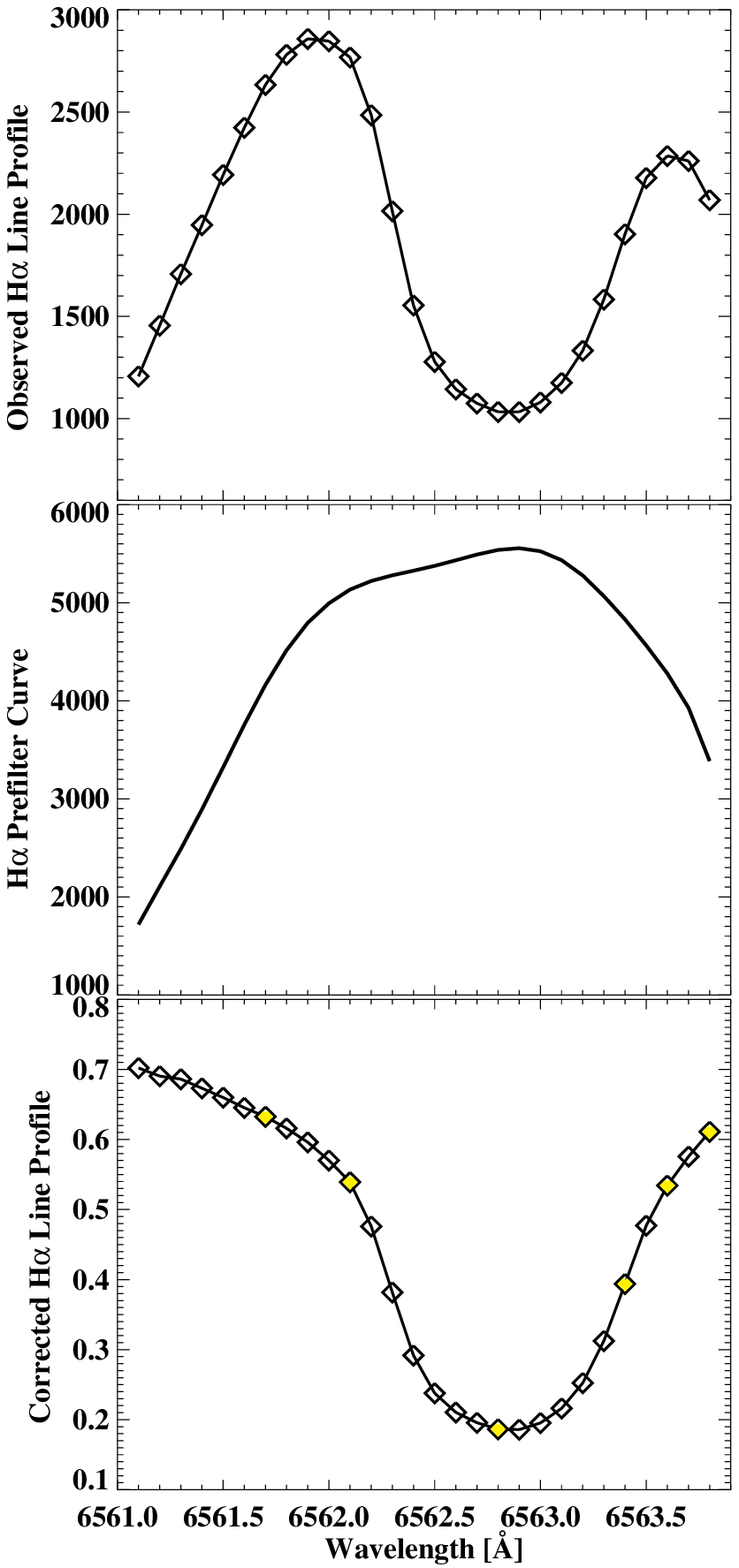}
  \caption{From the top down: An observed \ha\ line profile with prefilter transmission curve superposed, the \ha\ prefilter transmission curve for our observation, and the corrected \ha\ line profile with prefilter transmission curve removed and normalized to the continuum intensity. The diamond symbols represent the IBIS spectral sampling of our observation. The yellow diamond symbols highlight several wavelengths that will be used to display the narrowband filtergrams in Figures~\ref{FIG:HaImgs}~and~\ref{FIG:3Cores}.}
  \label{FIG:Pref}
\end{figure}

To obtain the prefilter transmission curve that is superimposed on all observed spectral profiles (see Figure~\ref{FIG:Pref}), we simply divide an average flat-field line profile by a well registered profile cropped from the Kitt Peak FTS (Fourier Transform Spectrometer) atlas of the disk center \citep{Neckel+Labs1984SoPh...90..205N, Neckel1999SoPh..184..421N} and smooth the resulting curve with a high degree polynomial fit. The flat-field spectral scans were taken at disk center with the telescope guider performing a random motion. The average flat-field line profile has been obtained from the calibrated flat-field spectra after averaging over all flat-field scans taken in about 10 minutes and subsequently correcting for the systematic wavelength shift across the FOV. Finally, all the observed spectra are corrected by dividing them with the acquired prefilter transmission curve. This procedure also naturally normalizes the observed \ha\ spectra to the continuum intensity as shown in Figure~\ref{FIG:Pref}.

We display the observed \ha\ spectra in two ways: original spectra $I(\lambda, t)$ and emission spectra $\Delta I(\lambda, t)=I(\lambda, t)-I(\lambda, t_0)$. The emission spectrum $\Delta I(\lambda, t)$ is defined as the difference between an enhanced line profile during a flare (i.e., the original spectra $I(\lambda, t)$) and a reference line profile $I(\lambda, t_0)$ that is averaged in the ribbon area but before the onset of the flare. This kind of difference spectra has been widely used to study the net emission component, especially for emissions that are not strong enough to flip over the line profile \citep[e.g.,][]{Canfield+etal1990ApJ...363..318C, Ding+F+H1995SoPh..158...81D, Johns-Krull+etal1997ApJS..112..221J}. The original spectra of optically thick chromospheric lines during flares are often complicated in shape and sometimes self-reversed. It is thus difficult and error-prone to deduce Doppler velocities from those original profiles using the bisector method \citep[e.g.,][]{Heinzel+etal1994SoPh..152..393H, Berlicki+etal2005A&A...430..679B, Berlicki+etal2006A&A...445.1127B}. Instead the net emission spectra are more regular in shape. Following previous studies \citep[e.g.,][]{Canfield+etal1990ApJ...363..318C, Ding+F+H1995SoPh..158...81D}, we derive bisectors (the middle points of horizontal chords intersecting the line profile at different levels) from the emission spectra to qualitatively illustrate the motion of \ha\ emitting materials.

\section{RESULTS AND ANALYSIS}\label{sec:results}

\subsection{The Circular Flare and its Remote Brightening}\label{sec:sdo}

\begin{figure*}[t]
  \epsscale{1.}
  \plotone{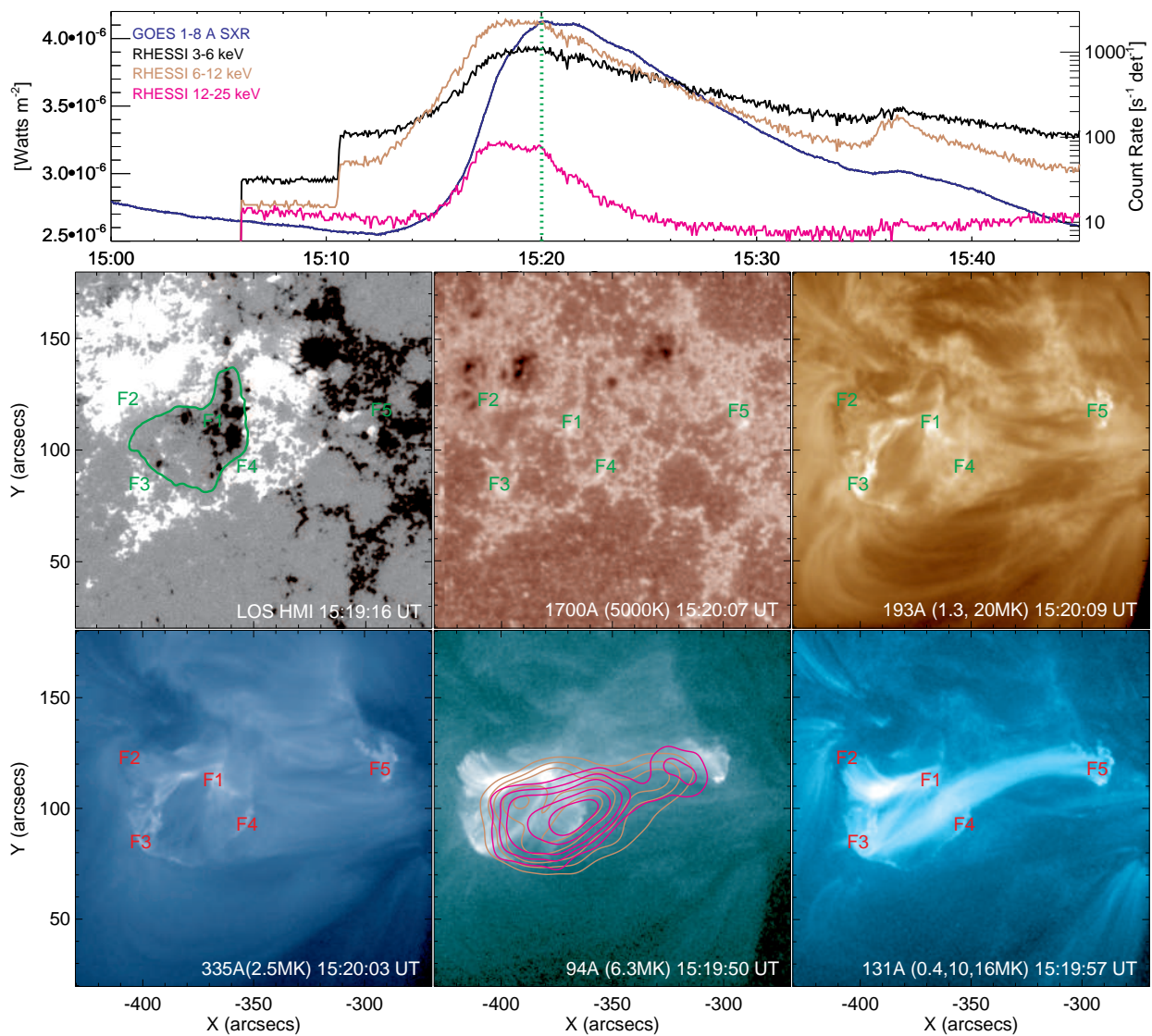}
  \caption{Upper Panel: GOES and RHESSI X-ray light curves of the C4.1 flare. Lower Panels: SDO HMI and AIA images show the magnetic field configuration and flare emissions at different temperatures (i.e., different atmospheric heights) around 15:20 UT when the HXR (RHESSI 12--25~keV) light curve has a second peak during the impulsive plateau. The RHESSI contours are plotted on the 94 \AA\ image. The green contour on the HMI image indicates the circular-like PIL of the parasitic magnetic configuration. F1--F5 denote the 5 flare footpoints in the low chromosphere. An animation of this figure is provided as online material to depict the complete evolution of the flare.}
  \label{FIG:Flare}
\end{figure*}

Figure~\ref{FIG:Flare} and the accompanying online movie depict the entire C4.1 flare that occurred in active region NOAA 11324 based on SDO observations. This confined flare shows five conspicuous footpoints (F1--F5) in the low chromosphere (AIA 1700 \AA\ image). The flare consists of a main part (circular-like ribbon) in the east (F1--F4) and a remote brightening in the west (F5). The circular-shaped ribbon is most prominent in the AIA 335 \AA\ image, where F1 sits inside the circle and fans out to the outer circle connecting F2--F4. In the magnetogram, F1--F4 form a so-called ``parasitic magnetic configuration''. The negative polarity F1 is encompassed by the positive polarities (F2--F4). Therefore a circular-like magnetic polarity inversion line (PIL) is naturally formed between F1 and F2--F4. F5 resides in a remote region outside of the parasitic configuration and shares the same polarity (i.e., negative polarity) as the parasitic F1. This kind of magnetic configuration seems to be a common feature for all observed circular flares involving 3D fan-spine reconnection \citep{Ugarte-Urra+W+W2007ApJ...662.1293U, Masson+etal2009ApJ...700..559M, Reid+etal2012A&A, Wang+Liu2012ApJ...760..101W, LiuC+etal2013ApJ}. The AIA 131 \AA\ image clearly shows that the remote footpoint (F5) is connected to the circular flaring fan by a long spine loop.

The RHESSI 12--25 keV light curve indicates that the C4.1 flare has two impulsive phases from 15:17:00 to 15:20:40~UT. The first and second impulsive phases peak around 15:18 and 15:20~UT, respectively. The left leg of F5 is brightened locally during the first impulsive phase then the right leg of F5 shows brightening and fast expansion during the second impulsive phase. Interestingly, the circular-ribbon flare studied by \citet{Masson+etal2009ApJ...700..559M} and \citet{Reid+etal2012A&A} also indicated two impulsive phases in the HXR light curve.

\subsection{IBIS \ha\ Imaging Spectroscopy of the Flare's Remote Ribbon}

\begin{figure*}[t]
  \epsscale{0.8}
  \plotone{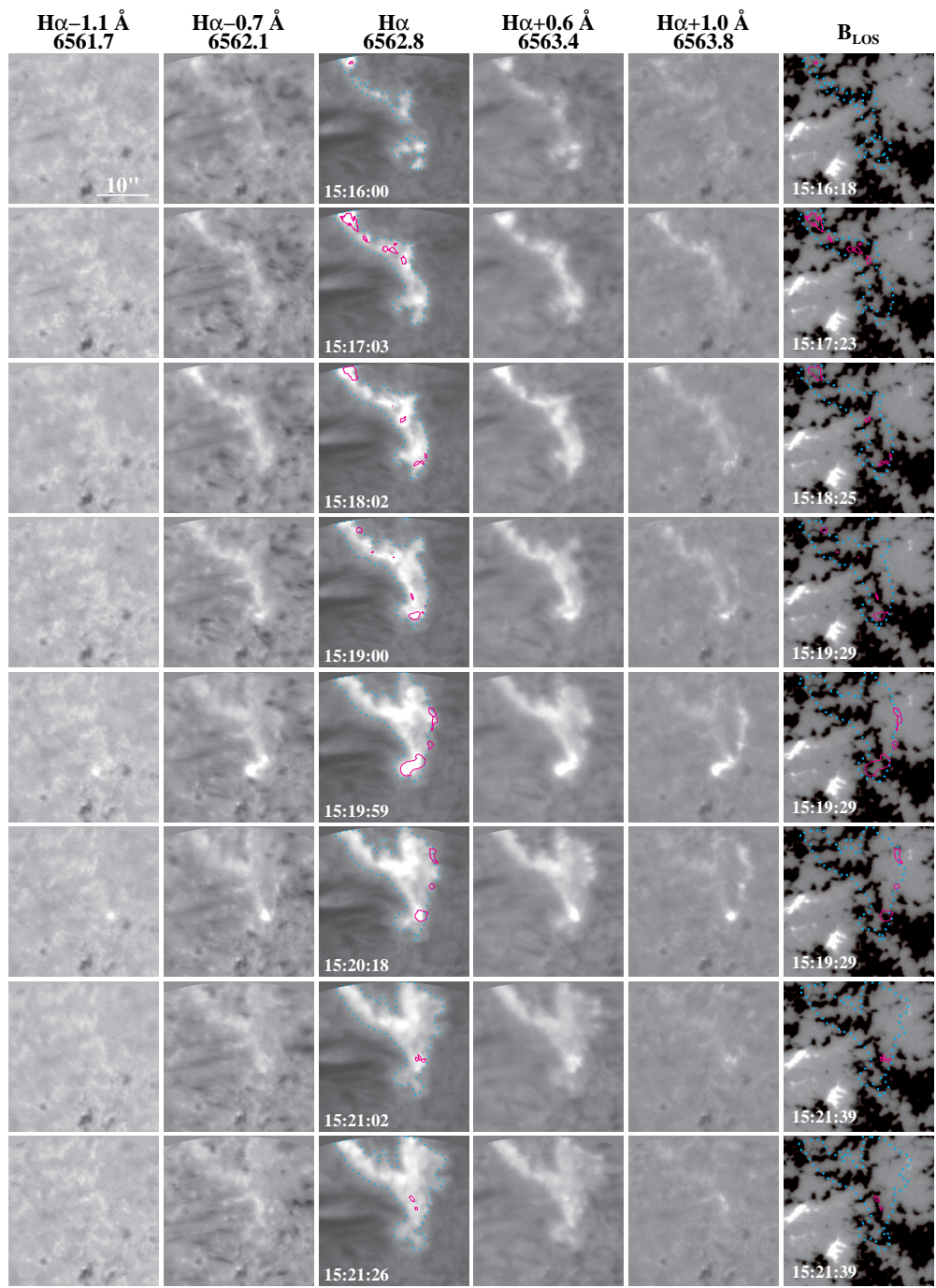}
  \caption{IBIS narrowband filtergrams show the evolution (top to bottom) of the flare remote ribbon (F5 in Figure~\ref{FIG:Flare}) at different wavelengths (left to right) crossing the \ha\ line. Only a 30$^{\prime\prime} \times$ 30$^{\prime\prime}$ sub-region in the upper-middle of the IBIS FOV is shown. The right-most column shows corresponding Hinode NFI Na\,\textsc{i}\,D$_1$ LOS magnetograms. The blue dotted contours encompass the entire brightening ribbon appearing in the \ha\ line center image (6562.8  \AA). The magenta solid contours outline the most powerful flare ``core'' structures that are obtained by selecting the brightest areas in the red wing image at 6563.8 \AA~(i.e., \ha$+1.0$~\AA). An animation is provided as online material to depict the evolution of the ribbon at multiple wavelengths across the \ha\ line profile. The wavelengthes are labeled on top of each narrowband image.}
  \label{FIG:HaImgs}
\end{figure*}

IBIS captured the remote ribbon (F5) of the circular flare and fully covered its temporal evolution. Figure~\ref{FIG:HaImgs} illustrates the impulsive-phase evolution of the brightening ribbon in narrowband filtergrams of five wavelengths spanning the \ha\ line. The ribbon lies in the upper-middle of the IBIS FOV and close to the AO lock point (the small pore in the lower-middle of the panels). The LOS magnetograms obtained by Hinode NFI provide the detailed magnetic field configuration. The entire ribbon (F5) resides in the negative polarity.

The ribbon area appears largest and has the highest contrast in the \ha\ line center images from which we outline the entire flaring area (blue contours) with an intensity threshold of 1.55 times the quiet region intensity. Toward the line wings that form at progressively deeper layers, the ribbon area and contrast decrease, finally only the most powerful flare kernels and ribbon fronts that penetrate to the deepest layer can be seen. We define these strongest flare kernels that survive in the observed reddest (6563.8 \AA, i.e., \ha$+1.0$\AA) wing images as ``core'' (magenta contours with a 20\% contrast enhancement in the 6563.8 \AA\ images) and other brightening areas as ``halo'' (i.e., the areas in between the blue and magenta contours). The cores that are the brightest in the line wing images are not necessarily the brightest in the line center images, see e.g., the 15:20:18~UT panels. The size of the compact cores is typically 1--3 $^{\prime\prime}$.

As mentioned in Section~\ref{sec:sdo}, the dynamic morphology of the entire ribbon and the motions of the ribbon cores show a distinct two-stage evolution during the two impulsive phases correspondingly. During the first impulsive phase from $\sim$15:16 to $\sim$15:19~UT, the ribbon was brightened locally in the left leg. The cores lie in the centroid of the ribbon and move parallel along the ribbon. During the second impulsive phase from $\sim$15:19 to $\sim$15:21~UT, the right leg of the ribbon brightens and expands quickly toward the west. The cores are mainly located at the expanding front and move perpendicular to the ribbon. A strong core in the southern part of the ribbon jumped between discrete magnetic elements with an apparent velocity of about 30 \kms\ during the second impulsive phase. This jumping strong core can even be seen in the bluest wing image observed (6561.1 \AA, i.e., \ha$-1.7$\AA). The speed of the ribbon's expansion toward the west is also about 30 \kms. The ribbon cores disappeared in the decay phase of the flare. Similar two-stage evolution of flare ribbons has also been observed and studied in details for typical two-ribbon flares \citep[e.g.,][]{Moore+etal2001ApJ...552..833M, QiuJ2009ApJ...692.1110Q}

\begin{figure*}[t]
  \epsscale{1.}
  \plotone{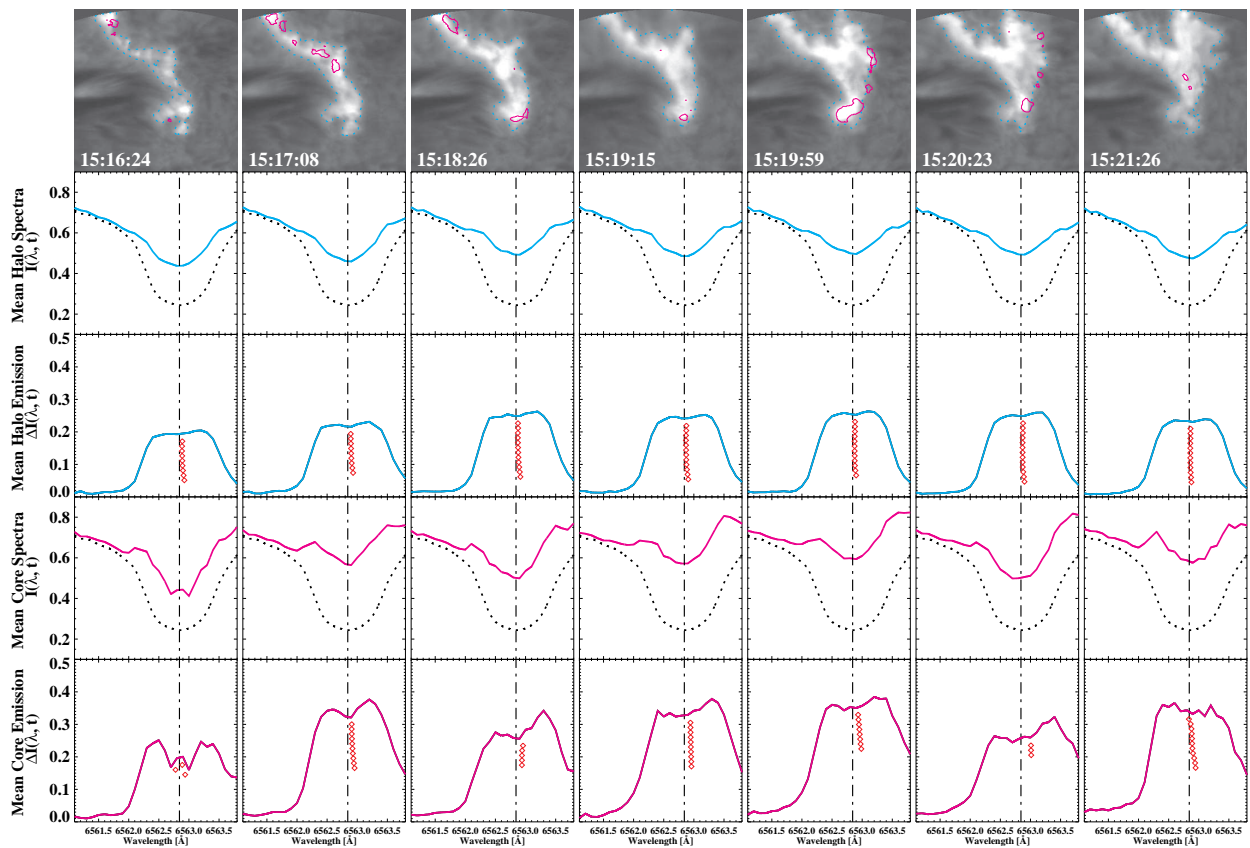}
  \caption{The \ha\ original and emission spectra averaged in the flare ribbon halo (blue curves) and core (magenta curves) areas at several times during the flare impulsive phase. The black dotted profiles plotted along with original enhanced spectra are the reference \ha\ line profile that is averaged over the ribbon area but before the flare. The difference between the original enhanced spectra due to flare and the reference spectrum gives the emission spectra. The vertical dash-dotted line marks the \ha\ line center at rest. The red diamond symbols are the bisectors of the emission spectra. The morphology of the ribbon at each time is shown on the top with \ha\ line center (6562.8 \AA) images.}
  \label{FIG:HaloCore}
\end{figure*}

Figure~\ref{FIG:HaloCore} displays the IBIS observed \ha\ spectra averaged over the flare ribbon halo and core areas in two ways: original spectra and emission spectra, at a few times during the impulsive phase. The flare causes enhanced \ha\ profiles which, however, are not strong enough to directly show the emission profile due to the small magnitude of the flare (C4.1). In this case, the defined emission spectra (i.e., the difference spectra between the flaring enhanced spectra and the reference pre-flare spectrum) become more useful to reveal the net emission. Comparing the emission spectra in halo and core areas, we find that those in core areas are stronger and wider. Moreover, the emission spectra in core areas show obvious central reversal signatures, which is not the case for those in halo areas. The bisectors of emission spectra in core areas show clear redshifts in contrast to those in halo areas. The redshifts of bisectors tend to increase from the line center to the line wing. The \ha\ emission is generally stronger in the red wing than in the blue wing, especially for the core areas, giving rise to the red asymmetry. Combining all the observed morphological, dynamical, and spectral characteristics shown in Figures~\ref{FIG:HaImgs} and~\ref{FIG:HaloCore}, it is reasonable to suggest that the cores represent the sites of intense nonthermal electron beam precipitation. Thus, the motion of cores tracks magnetic reconnection processes involving different flux tubes. In contrast, the heating in the halo areas is most likely due to other mechanisms, such as thermal conduction and SXR/EUV radiation. Such a core-halo structure of flare ribbons was also found in \citet{Neidig+etal1993ApJ...406..306N} and \citet{XuY+etal2006ApJ...641.1210X} for white-light flares based on continuum or \ha\ wing images. The authors used direct heating by nonthermal particle beam and indirect heating such as chromospheric back-warming to interpret the white-light emissions in the core and halo areas, respectively.

\begin{figure*}[t]
  \epsscale{0.95}
  \plotone{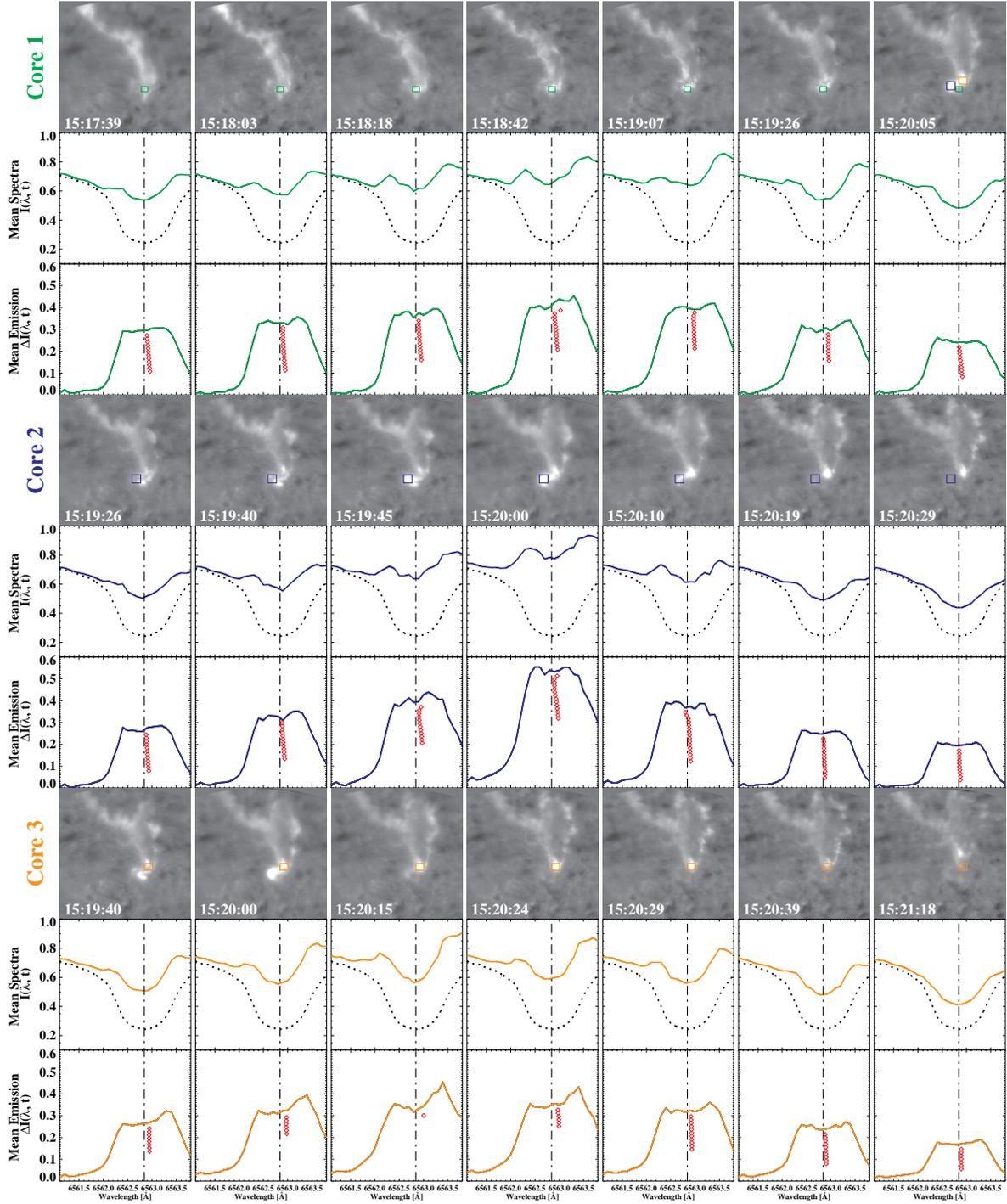}
  \caption{The temporal evolution of \ha\ original and emission spectra averaged in three fixed positions (the colored boxes) where a strong core passes by. We simply name the three core positions as Core 1, 2, and 3. The red wing images at 6563.6~\AA~(i.e., \ha$+0.8$\AA) are shown accordingly. The upper right panel illustrates the context of the three core positions. The black dotted profiles are the reference pre-flare spectrum. The red diamond symbols are the bisectors of the emission spectra.}
  \label{FIG:3Cores}
\end{figure*}

Examining the temporal evolution of \ha\ spectra at a fixed position where a core once appeared can provide clues on the heating and cooling processes of the chromosphere which is directly bombarded by energetic electrons. Figure~\ref{FIG:3Cores} illustrates how the local spectra evolve at three fixed positions when a strong core moves into and leaves from the positions. The local emission increases as the core appears then decreases as the core disappears in those locations. Only when the core lies in the position, the emission becomes wider and centrally reversed. The redshifts of emission spectra bisectors also tend to be largest then. During the core's existence, the emission in the red wing becomes stronger than that in the blue wing, which is more conspicuous in the original spectra. The excess red-wing emission gives rise to the red asymmetry. All these spectral characteristics and their temporal evolution are consistent with previous studies of electron beam precipitation sites where the intense heating of the chromosphere by nonthermal electrons causes explosive evaporation in the upper chromosphere and consequently downward moving condensation in the lower chromosphere \citep[e.g.,][]{Ichimoto+Kurokawa1984SoPh...93..105I, Canfield+Gunkler1985ApJ...288..353C, Canfield+etal1990ApJ...363..318C, Canfield+etal1993ApJ...411..362C, LeeS+etal1996ApJ...470L..65L}.

\begin{figure*}[t]
  \epsscale{1.}
  \plotone{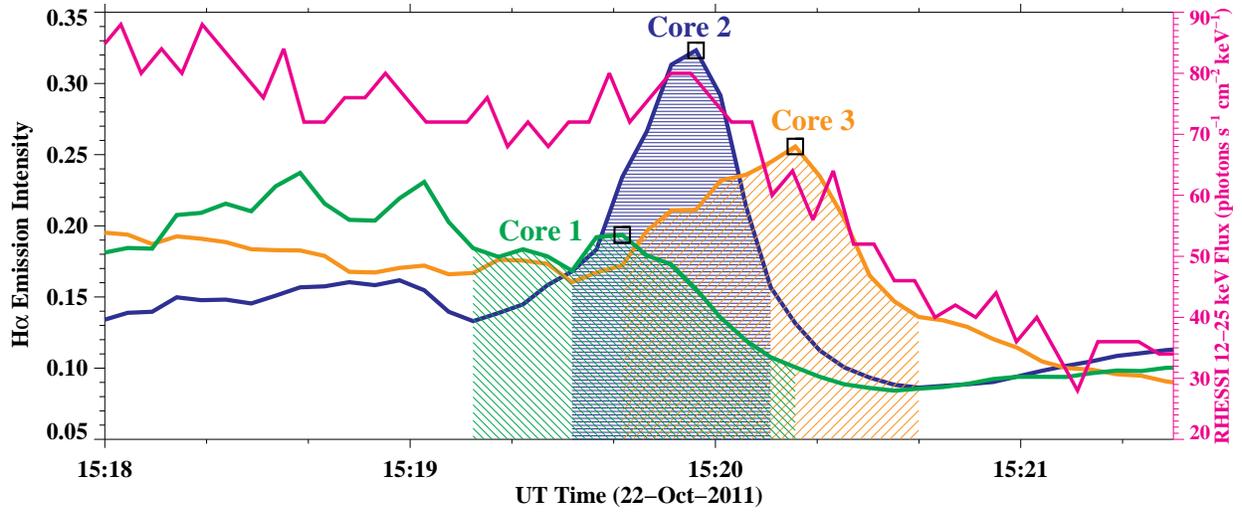}
  \caption{The light curves of \ha\ emission intensity in the three core positions defined in Figure~\ref{FIG:3Cores} showing the heating and cooling time profiles. The emission intensity is averaged over the entire \ha\ line observed. As comparison, the RHESSI HXR (12--25 keV) light curve is also plotted in magenta.}
  \label{FIG:Cool}
\end{figure*}

The light curves of \ha\ emission in the three core positions are plotted in Figure~\ref{FIG:Cool} as a function of time. They illustrate the heating and cooling time profile of the regions bombarded by energetic electrons. The maxima of these light curves prior to fast cool-down are denoted with black boxes. It is interesting that these maxima occurred at different locations correlate well with the sub-peaks of the RHESSI HXR light curve with only a slight delay. This implies that the sub-bursts of HXR may correspond to  different locations. The multiple peaks at the location of core 1 suggest that one location can also be bombarded for several episodes. Amongst the three core locations, the location of core 2 shows the fastest heating and cooling time profile, because the core moved quickly and stayed there for only short time (less than 30~s). In contrast, the core persisted in location 1 for more than one minute, so the heating timescale is not clear for the location of core 1. From the light curves (the shaded areas), we estimate that the heating times for the locations of core 2 and 3 are about 24 and 33~s, respectively. The cooling times (the time taken for the emission intensity to decrease to half of the maximum value) for the three core positions are about 33, 14, and 24~s, respectively.

\subsection{Comparison between \ha\ and HXR Emission}
\begin{figure*}[t]
  \epsscale{0.8}
  \plotone{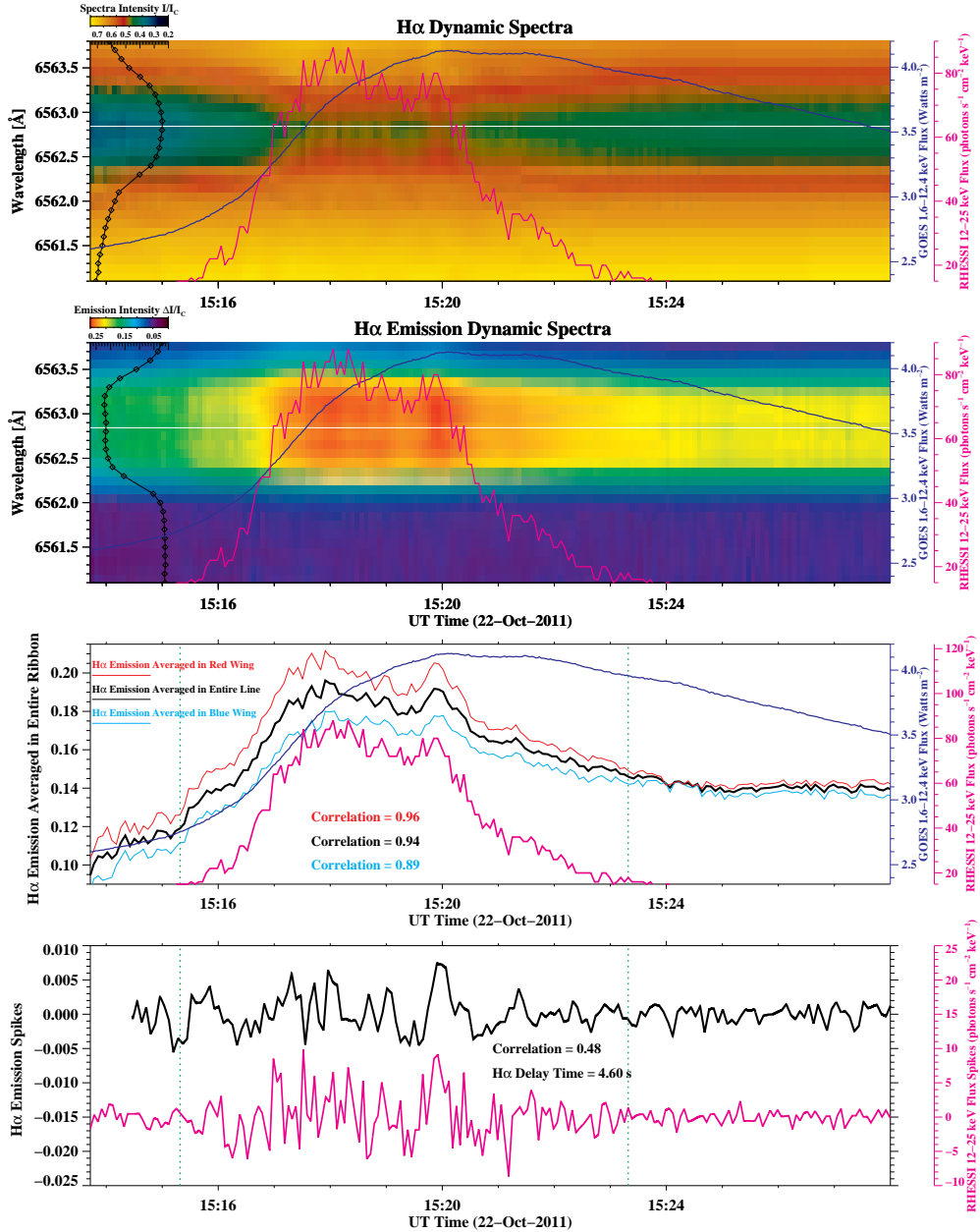}
  \caption{Top two panels: \ha\ dynamic spectra and emission dynamic spectra, i.e., color-coded \ha\ original or emission spectrum (along the $y$-axis) as a function of time. Each spectrum is spatially averaged over the entire brightening ribbon. The \ha\ line center at rest is marked with a white line. The GOES SXR and RHESSI HXR light curves are overplotted as purple and magenta curves. 3rd panel: light curves of \ha\ emission averaged symmetrically over the red wing, the entire line core, and the blue wing are compared with GOES and RHESSI light curves. The correlation coefficients between \ha\ and RHESSI HXR light curves within the impulsive time interval (inbetween green vertical lines) are labeled. 4th panel: the light curves of \ha\ emission averaged over the entire line core (black) and RHESSI HXR (magenta) are subtracted by their overall evolution curves, so only emission spikes are left and compared.}
  \label{FIG:HaHXR}
\end{figure*}

As shown in the top two panels of Figure~\ref{FIG:HaHXR}, we construct dynamic spectra using a time series of IBIS observed \ha\ original and emission profiles that are spatially averaged over the entire brightening ribbon. Both types of dynamic spectra show two major \ha\ emission phases that are cotemporal with the two impulsive phases of the HXR light curve. Embedded in the two major impulsive phases, several episodes of \ha\ emission can be seen that seem to correlate with the sub-peaks of HXR emission.

For a quantitative comparison between \ha\ and HXR emission, we compute the time profiles of the \ha\ emission intensity based on the dynamic emission spectra. The \ha\ emission spectrum at each time is averaged symmetrically over the red wing (6562.8--6563.8 \AA), the entire line core (6561.8--6563.8 \AA), and the blue wing (6561.8--6562.8 \AA), giving the light curves for the three wavelength bands shown in the 3rd panel. These light curves clearly show that the \ha\ emission in the red wing exceeds that in the blue wing during the impulsive phase by about 15\%. Other than that, the temporal variations for the three wavelength bands are quite similar. The \ha\ emission light curves closely resemble that of the HXR emission in the rising and impulsive phases, but decay slower during the gradual (decaying) phase, where they resemble more that of the SXR emission that originates from hot coronal loops. This kind of behavior is consistent with the known characteristics of flare optical emissions \citep[e.g.,][]{Zirin+Tang1990ApJS...73..111Z}. It also implies that in the impulsive phase the \ha\ emission is dominated by nonthermal electron beam heating, while in the gradual phase it is mainly due to coronal thermal conduction and/or radiation. Our observation shows a very high temporal correlation ($\gtrsim0.9$) between \ha\ and HXR emission during the impulsive phase, with higher correlation in the red wing emission than in the blue wing emission. Note that the RHESSI HXR light curve is measured from the entire flaring region, while the \ha\ emission is only measured from the remote ribbon of the flare. Their close temporal correlation implies that the remote brightening is actively involved in the 3D reconnection process and directly heated by reconnection-accelerated precipitating electron beams.

Both HXR and \ha\ emission light curves show short timescale fluctuations (spikes or episodes) superposed on the general evolution curve. These fast fluctuations are likely signatures of elementary reconnection and energy transfer processes. Seeking the correlation between the HXR and \ha\ emission fluctuations can provide clues on their emission timings and the chromospheric heating mechanism during a flare \citep{WangH+etal2000ApJ...542.1080W}. The 4th panel of Figure~\ref{FIG:HaHXR} illustrates the fluctuations of \ha\ and HXR emission by removing their general evolution curves. Shifting the \ha\ fluctuations back and forth, we find that the correlations between \ha\ and HXR fluctuations reach a unique maximum value (linear correlation = 0.48) when the \ha\ fluctuations curve is shifted forward by 4.6~s. This provides a quantitative assessment of the time delay of \ha\ emission relative to HXR emission. At the current temporal resolution (HXR 4~s, \ha\ 4.8~s), the estimated time delay is 4.6~s for the observed remote flare ribbon. Higher temporal resolution could increase the statistic significance of this kind of measurements. The time delays in \ha\ emission with respect to HXR emission have been statistically studied by \citet{Radziszewski+R+P2011A&A...535A.123R} for many flare kernels and their sub-bursts. The authors found two distinct groups of time delays. The short delay (1--6 s) group is associated with fast heating by nonthermal electrons, while the longer delay (10--18 s) group is ascribed to a slower energy transfer mechanism such as a moving conduction front \citep{Trottet+etal2000A&A...356.1067T}. Our observed 4.6~s delay falls into the short delay group, thus reinforcing the electron beam heating mechanism for the remote brightening during the impulsive phase. \citet{WangH+etal2000ApJ...542.1080W} attributed their observed 2--3~s time delay between the \ha\ blue wing and HXR bursts in the early impulsive phase of a C5.7 flare to the ionization timescale of the cool chromosphere \citep{Canfield+Gayley1987ApJ...322..999C}. The time lags due to temperature structure evolution, hydrogen ionization and emission timescales, and energy deposit rate as proposed by \citet{Kasparova+etal2009A&A...499..923K} may explain the delay of \ha\ bursts relative to HXR bursts in our observation. In addition, the long spine loop for the electrons to travel from the 3D null point reconnection site (most probably somewhere in the corona above the main circular ribbon) to this remote chromospheric precipitation site may also contribute to the time delay, since the majority of HXR emission comes from the main circular ribbon site. We estimate the spine loop length to be about 100$^{\prime\prime}$ from Figure~\ref{FIG:Flare}. The speed of nonthermal electrons of 17~keV is about a quarter of the speed of light. So it will take those electrons about 1~s to travel along the spine loop.

\section{SUMMARY AND PERSPECTIVE}\label{sec:summary}

We present unprecedented 2D \ha\ imaging spectroscopy of a circular flare's remote ribbon observed by IBIS with high spatial (0.1\arcpix\ image scale), temporal (4.8~s cadence), and spectral (0.1~\AA\ stepsize) resolution. The main circular ribbon part of the flare occurred in a parasitic magnetic configuration, which seems to be a common feature for all observed circular-ribbon flares accompanied by a jet and/or a remote brightening involving a 3D fan-spine magnetic topology.

We defined ``core'' and ``halo'' structures in the impulsive-phase flare ribbon based on narrowband images in the \ha\ line wing and line center. Only the emission spectra in the core areas showed typical spectral characteristics of theoretically simulated \ha\ line profiles of flaring chromosphere heated by nonthermal electron beam precipitation. These spectral characteristics include: broad and centrally reversed emission spectra, excess emission in the red wing than in the blue wing (i.e., red asymmetry), and redshifted bisectors of the emission spectra. In addition, we observed rather quick heating ($\sim$30~s) and cooling ($\sim$14--33~s) timescales in the flare core positions, suggesting impulsive heating of the core areas by nonthermal electrons. Therefore, the flare cores revealed by IBIS track the sites of electron beam precipitation and reflect magnetic reconnection processes with exceptional spatial and temporal resolution. The flare cores showed two-stage motion during respectively the two impulsive phases of the flare. The cores first move parallel along the ribbon for about 3 minutes during the first impulsive phase and then move perpendicular to the ribbon causing fast expansion of the ribbon during the second impulsive phase. Some strong cores jump quickly (30~\kms) between discrete magnetic elements implying reconnection involving different flux tubes. The cores disappeared in the decay phase of the flare. The heating of the halo is more likely due to thermal conduction and/or SXR/EUV radiation.

Our observation shows a very high temporal correlation ($\gtrsim0.9$) between the integrated \ha\ and HXR emission light curves during the flare impulsive phase, with higher correlation in the red wing emission than in the blue wing emission. Further comparison of the sub-bursts between the \ha\ and HXR emission light curves revealed a 4.6~s time delay in the \ha\ emission compared to the HXR emission. The short time delay reinforces the electron beam heating mechanism for the remote ribbon of the circular flare during the impulsive phase. The ionization timescale of the cool chromosphere and the long spine loop for the electrons to travel may contribute to the short delay in the \ha\ emission of the remote flare ribbon with respect to the HXR emission of the entire flare. The sub-bursts of HXR/\ha\ emission may correspond to different locations. One location can also be bombarded for several episodes.

Besides IBIS, other FPI-based 2D imaging spectrometers are either being operated or commissioned around the world, such as the Telecentric Etalon SOlar Spectrometer \citep[TESOS;][]{Kentischer+TESOS1998A&A...340..569K, Tritschler+TESOS2002SoPh..211...17T} at the 75~cm Vacuum Tower Telescope (VTT) in Tenerife, the CRisp Imaging SpectroPolarimeter \citep[CRISP;][]{Scharmer+etal2008ApJ...689L..69S} at the 1~m Swedish Solar Telescope (SST), the GREGOR Fabry-P\'erot Interferometer \citep[GFPI;][]{Denker+GFPI2010SPIE.7735E.217D, Puschmann+etal2012AN....333..880P} at the 1.5~m GREGOR telescope in Canary Islands, and the Fabry-P\'erot system at the 1.6~m New Solar Telescope (NST) in Big Bear Solar Observatory (BBSO). Moreover, a spectral imager based on two FPIs for studying the lower solar atmosphere is planned at the focal plane of the 50~cm telescope onboard the ADAHELI (ADvanced Astronomy for HELIophysics) mission, an Italian Space Agency small satellite mission \citep{Berrilli+etal2010AdSpR..45.1191B}. All these instruments are anticipated to provide valuable spectroscopy or spectropolarimetry with high spatial, spectral, and temporal resolution for scrutinizing flares and other highly dynamic or small-scale phenomena on the Sun.

\acknowledgments
This work was supported by NASA under grants NNX08AQ32G, NNX11AQ55G, NNX13AF76G \& NNX13AG13G and by NSF under grants AGS 0936665, 1153226, 0839216 \& 1153424. CD was supported by grant DE 787/3-1 of the Deutsche Forschungsgemeinschaft (DFG). We thank Dr. Etienne Pariat for carefully reading the paper and providing valuable comments. IBIS is a project of INAF/OAA with additional contributions from Univ. of Florence and Rome and NSO. We thank the NSO/SP observers (Doug Gilliam, Joe Elrod, and Mike Bradford) for their professional and excellent observing support. We also appreciate the great support from Hinode and communications with Dr. Toshifumi Shimizu and other Hinode team members before and during the successful IBIS-Hinode coordinated observing campaign. The National Solar Observatory is operated by the Association of Universities for Research in Astronomy under a cooperative agreement with the National Science Foundation, for the benefit of the astronomical community. Hinode is a Japanese mission developed and launched by ISAS/JAXA, with NAOJ as domestic partner and NASA and STFC (UK) as international partners. It is operated by these agencies in co-operation with ESA and NSC (Norway). This research has made use of NASA's Astrophysics Data System (ADS).

\end{document}